\documentstyle[12pt]{article}
\begin{document}
\hbadness=10000
\hbadness=10000
\begin{titlepage}
\nopagebreak
\begin{flushright}
{\normalsize
HIP-1998-55/TH\\
August, 1998}\\
\end{flushright}
\vspace{0.7cm}
\begin{center}

{\large \bf  Vacuum Structure of Softly Broken $N=1$ Supersymmetric QCD}

\vspace{1cm}

{\bf Masud Chaichian  
\footnote[1]{e-mail: chaichia@rock.helsinki.fi}} 
and 
{\bf Tatsuo Kobayashi 
\footnote[2]{e-mail: kobayash@rock.helsinki.fi}}\\

\vspace{0.7cm}
Department of Physics, High Energy Physics Division\\
    University of Helsinki \\
and\\
Helsinki Institute of Physics\\
    P.O. Box 9 (Siltavuorenpenger 20 C)\\
    FIN-00014 Helsinki, Finland \\

\end{center}
\vspace{0.7cm}

\nopagebreak

\begin{abstract}
We study softly broken $N=1$ supersymmetric QCD with 
the gauge group $SU(N_c)$ and $N_f$ flavors of quark pairs.
We investigate vacuum structure of the theory with generic 
soft supersymmetry breaking terms.
Trilinear soft breaking terms play an essential role in determining vacua.
For $N_f=N_c+1$, chiral symmetry is broken for a sufficiently large magnitude 
of trilinear couplings, while it is unbroken in the case with only soft masses.
In the case where appearance of trilinear coupling terms is allowed, i.e. 
for $N_f \geq N_c+1$, 
we have two possible vacua, the trivial and non-trivial ones.
Otherwise, we only have the non-trivial vacuum, which corresponds to 
the non-trivial vacuum in the $N_f \geq N_c+1$ theory.
\end{abstract}
\vfill
\end{titlepage}
\pagestyle{plain}
\newpage
\def\thefootnote{\fnsymbol{footnote}}

\section{Introduction}
Recently, nonperturbative aspects of $N=1$ supersymmetric (SUSY) QCD have 
been understood \cite{seiberg,rev}.
It is very important to extend such analyses to non-SUSY QCD 
and study its nonperturbative aspects, confinement and chiral symmetry 
breaking.
To this end, it is an interesting trial to study softly broken 
$N=1$ SUSY QCD.
Actually in Refs.\cite{soft1,soft2,soft3} $N=1$ SUSY QCD 
with soft scalar masses as well as gaugino masses 
has been discussed and interesting 
results have been obtained.
Also softly broken $N=2$ SUSY QCD has been studied in Ref.\cite{soft4}. 
In addition, non-supersymmetric QCD is recently discussed from the 
viewpoint of brane dynamics \cite{mqcd}.

In particular, the vacuum structure of $N=1$ SUSY QCD broken by 
adding soft masses has been clarified for the theory with $SU(N_c)$ 
gauge group and $N_f$ flavors of quark pairs in Ref.\cite{soft1}.
For $N_f<N_c$, there is a nontrivial stable vacuum, while there is no 
vacuum in the SUSY limit \cite{seiberg}.
For $N_f=N_c$ we can have two nontrivial vacua and there is no trivial 
vacuum as in the SUSY limit.
In one vacuum, only the meson fields $T$ develop their expectation values 
(VEVs)  and in the other only the baryon fields $B$ and $\bar B$ develop 
their VEVs.
Which vacuum is realized depends on the soft mass ratio between $T$ and 
$B$ ($\bar B$).
In both vacua chiral symmetry is broken and this situation is 
the same as the SUSY limit, where we have chiral symmetry breaking as well as 
confinement.
On the other hand, for $N_f=N_c+1$ we have only the trivial vacuum and 
chiral symmetry is not broken, while in the SUSY limit we have confinement 
without chiral symmetry breaking, i.e. s-confinement.
For $N_f > N_c+1$, we have only the trivial vacuum and the presence of 
the Seiberg duality is suggested even in SUSY QCD broken by soft mass terms.
Furthermore, an attempt to relate soft masses between dual theories 
has been done in Ref.\cite{soft5}.

For our purpose it is useful to add all the allowed soft SUSY breaking terms.
Because we do not know which region of the whole soft SUSY breaking 
parameter space corresponds to the real non-SUSY QCD.
However, we expect that generic study on softly broken $N=1$ SUSY QCD, 
i.e. with generic soft SUSY breaking terms, can present the real 
non-SUSY QCD.
For that purpose, in Ref.\cite{soft6} softly broken SUSY QCD with generic 
soft SUSY breaking terms has been discussed for $N_f > N_c+1$.
It has been shown that trilinear soft SUSY breaking terms are important 
to determine a vacuum in the dual theory and we have found the nontrivial 
vacuum for a sufficiently large magnitude of the trilinear coupling.
Furthermore, the presence of duality between non-SUSY theories has been 
suggested even in the broken phase.

In this paper, we extend such analyses to the case $N_f \leq N_c+1$, 
which might correspond more to a real world with the $SU(3)$ colour 
symmetry below the GeV scale when studying the confinement region 
with an effective $N_f$ without heavy quarks.
Moreover, in this case we obtain confinement in the SUSY limit.
We investigate the vacuum structure of softly broken SUSY QCD with 
generic soft SUSY breaking terms.
Further, we study relations among non-trivial vacua corresponding to 
different flavour numbers.

This paper is organized as follows.
In the next section, we review the vacuum structure of softly 
broken $N=1$ SQCD for $N_f > N_c+1$.
Also we give a brief review on deformation of the flavour number in 
the SUSY limit.
In section 3, we study the vacuum structure of the case with 
$N_f=N_c+1$ and show that the nontrivial vacuum leading to chiral symmetry 
breaking can be realized beside the trivial vacuum with chiral symmetry.
In section 4, we study the vacuum structure of the case with 
$N_f = N_c$.
We show the vacuum structure is the same as the case with only soft masses 
added, that is, we find only the nontrivial vacuum.
We consider the nontrivial vacuum for $N_f \leq N_c$ relating it to 
the vacuum structure for $N_f=N_c+1$.
Section 5 is devoted to conclusions and discussions.

\section{$N_f> N_c+1$}

At first we give a brief review of softly broken 
$N=1$ supersymmetric QCD for $N_f> N_c+1$.
We concentrate on the case with $N_c \geq 3$.
We consider the $N=1$ supersymmetric QCD with the gauge symmetry $SU(N_c)$ 
and $N_f$ flavours of quark supermultiplets, $\widehat Q^i$ and 
$\widehat {\overline Q}_i$.
This theory has the flavour symmetry 
$SU(N_f)_Q\times SU(N_f)_{\overline Q}$ and no superpotential.
In the case with $N_f> N_c+1$, the dual theory is described by 
the $N=1$ SUSY theory with the gauge group $SU(N_f-N_c)$, 
$N_f$ flavours of dual quark pairs $\widehat q_i$ and 
$\widehat {\overline q}^i$, and 
$N_f\times N_f$ singlet meson supermultiplets ${\widehat T}^i_j$.
The dual theory has the same flavour symmetry as the electric theory and 
the dual theory has the superpotential,
\begin{eqnarray}
W=\widehat q_i {\widehat T}^i_j\widehat {\overline q}^j.
\label{spot}
\end{eqnarray}

In the dual theory, all the symmetries except $R$-symmetry allow the 
following soft SUSY breaking terms,
\begin{eqnarray}
{\cal L}_{SB} &=& -m_q^2 {\rm tr} |q|^2-m_{\bar q}^2 {\rm tr} 
|\overline q|^2 - 
m_T^2 {\rm tr} |T|^2+(hq_iT^i_j\overline q^j + h.c.),
\label{ssb}
\end{eqnarray}
where $q_i$, $\overline q^i$ and $T^i_j$ denote scalar components of 
$\widehat q_i$, $\widehat {\overline q}^i$ and ${\widehat T}^i_j$, 
respectively.
Also the gaugino mass terms are added.
For the kinetic term, we assume the canonical form with 
normalization factors $k_q$ and $k_T$ for $q$, $\bar q$ and $T$.
Then we write the following scalar potential:
\begin{eqnarray}
V(q,\bar q,T) &=& {1 \over k_T}{\rm tr} (q q^\dagger \bar q^\dagger \bar q)
+{1 \over k_q}{\rm tr}(qTT^\dagger q^\dagger
+\bar q^\dagger T^\dagger T \bar q) \nonumber \\
&+& {\tilde g^2 \over 2}({\rm tr} q^\dagger \tilde t^a q -  
{\rm tr} \bar q \tilde t^a \bar q^\dagger)^2 
+m_q^2{\rm tr} q^\dagger q + m_{\bar q}^2 {\rm tr} \bar q \bar q^\dagger  
\nonumber \\ 
&+& m_T^2 {\rm tr} T^\dagger T - (hq_iT^i_j\bar q^j + h.c.),
\label{pot0}
\end{eqnarray}
where the third term is the $D$-term and $\tilde g$ denotes 
the gauge coupling constant of the dual theory.
In Ref.\cite{soft6} it has been shown that the trilinear coupling terms 
$hqT\overline q$ play a crucial role in determining the minimum 
of the potential.

We assume $h$ is real.
The minimum of potential can be obtained along the 
following diagonal direction ,
\begin{equation}
q=\left( \begin{array}{ccccc}
q_{(1)} &  & & 0 & \\
 &  q_{(2)} & & & \\
 0 &  & \cdots &  & \\
 & & &   q_{(\tilde N_c)} & 
   \end{array}\right),
\end{equation}
\begin{equation}
\bar q=\left( \begin{array}{cccc}
\bar q_{(1)} &  & & 0 \\
 &  \bar q_{(2)} & & \\
0 &  & \cdots &  \\
 & & &   \bar q_{(\tilde N_c)} \\
 & & & 
   \end{array}\right),
\end{equation}
\begin{equation}
T=\left( \begin{array}{ccccc}
T_{(1)} & & &  0 & \\
 &  T_{(2)} & & & \\
 0 &  & \cdots &  & \\
 & & & T_{(\tilde N_c)} & \\
 & & & & 0 
   \end{array}\right),
\label{Tdiag}
\end{equation}
where all the entries, $q_{(i)}$, $\bar q_{(i)}$ and $T_{(i)}$, 
can be made real.
Along the $D$-flat direction,
\begin{equation}
q_{(i)}=\bar q_{(i)}=X_i,
\end{equation}
the potential is written as 
\begin{eqnarray}
V(X,T) &=& \sum^{\tilde N_c}_{i=1}[{1 \over k_T}X_i^4+
(m_q^2+m_{\bar q}^2)X_i^2+m_T^2T_{(i)}^2 \nonumber \\
&+& {2 \over k_q}T_{(i)}^2X_i^2 - 2hT_{(i)}X_i^2].
\end{eqnarray}

This potential always has the nontrivial vacuum with $X_i \neq 0$ and 
$T_{(i)}\neq 0$ if 
\begin{eqnarray}
h \gg {m^2_q+m^2_{\overline q}  \over 2k_q}.
\end{eqnarray}
In addition, we have the nontrivial vacuum with $X_i \neq 0$ and 
$T_{(i)}\neq 0$ for a certain region of 
intermediate values of $h$, if 
$m^2_T/k_T$ is sufficiently large compared with 
$(m^2_q+m^2_{\overline q})/ 2k_q$ \cite{soft6}.
Otherwise, in particular for a sufficiently small value of $h$ we have 
the trivial vacuum with $T=0$ and $q=0$.

Next we give a brief review on deformation of the flavour 
number \cite{seiberg}.
We add a mass term of one flavour, e.g. 
$m\widehat Q^{N_f}\widehat {\overline Q}_{N_f}$, in 
the superpotential of the $N=1$ supersymmetric QCD, so that 
one flavour becomes massive and the flavour symmetry breaks into 
$SU(N_f-1)_Q \times SU(N_f-1)_{\overline Q}$.
That corresponds to add the term $m\widehat T^{N_f}_{N_f}$ in the 
superpotenial of the dual theory, i.e.
\begin{eqnarray}
W=\widehat q_i {\widehat T}^i_j\widehat {\overline q}^j + 
m\widehat T^{N_f}_{N_f}.
\end{eqnarray}
If we add only soft scalar masses, we have the following scalar potential, 
\begin{eqnarray}
V(q,\bar q,T) &=& {1 \over k_T}\sum_{i,j} |q_i \bar q^j
+\delta_i^{N_f}\delta^j_{N_f}m|^2
+{1 \over k_q}{\rm tr}(qTT^\dagger q^\dagger
+\bar q^\dagger T^\dagger T \bar q) \nonumber \\
&+& {\tilde g^2 \over 2}({\rm tr} q^\dagger \tilde t^a q -  
{\rm tr} \bar q \tilde t^a \bar q^\dagger)^2 
+m_q^2{\rm tr} q^\dagger q + m_{\bar q}^2 {\rm tr} \bar q \bar q^\dagger  
\nonumber \\ 
&+& m_T^2 {\rm tr} T^\dagger T.
\label{pot02}
\end{eqnarray}
In the SUSY limit the dual squark pair develops its VEV, 
\begin{eqnarray}
q^{N_f}\overline q_{N_f}=-m.
\end{eqnarray}
Therefore, the gauge symmetry and the flavour symmetry break into 
$SU(N_f-N_c-1)$ and $SU(N_f-1)_q \times SU(N_f-1)_{\overline q}$.
This breaking takes place even in the case with nonvanishing SUSY breaking 
parameters if SUSY breaking parameters are small enough compared with 
the mass $m$.
Hence, it is obvious that through this breaking the trilinear 
coupling $h$ in the $N_f$ flavour theory corresponds to one 
in the $(N_f-1)$ flavour theory.

Now we consider the $N_f$ flavour theory which has 
the superpotential (\ref{spot}) and the trilinear SUSY breaking term 
only for the $N_f$-th flavour $hq_{N_f}T^{N_f}_{N_f}\overline q^{N_f}$ 
as well as soft scalar mass terms.
Its scalar potential is the same as eq.(\ref{pot0}) except for replacing 
$q_iT^i_j\overline q^j$
by $q_{N_f}T^{N_f}_{N_f}\overline q^{N_f}$.
Thus, if $h$ is large enough, the gauge and flavour symmetry are broken as 
$SU(N_c) \rightarrow SU(N_c-1)$ and $SU(N_f) \rightarrow SU(N_f-1)$.
Let us compare such scalar potential with the scalar potential 
(\ref{pot02}).
If we fix $T^{N_f}_{N_f}$, we find that the term 
$hq_{N_f}T^{N_f}_{N_f}\overline q^{N_f}$ works effectively in a way similar 
to the mass term $mT^{N_f}_{N_f}$ in the scalar potential (\ref{pot02}).

\section{$N_f=N_c+1$}

For $N_f=N_c+1$, the $N=1$ supersymmetric QCD is described in terms of 
$(N_c+1) \times (N_c+1)$ meson fields $\widehat T^i_j$ and $(N_c+1)$ baryon 
fields $\widehat B_i$ and $\widehat {\overline B}^i$.
They have the superpotential,
\begin{eqnarray}
W={1 \over \Lambda^{2N_c-1}}
(\widehat B_i \widehat T^i_j\widehat {\overline B}^j -\det \widehat T).
\label{spot2}
\end{eqnarray}

The flavour symmetry allows the SUSY breaking trilinear coupling 
\begin{eqnarray}
h'B_i T^i_j{\overline B}^j.
\end{eqnarray}
Thus we consider here the following SUSY breaking terms,
\begin{eqnarray}
{\cal L}_{SB} &=& -m_B^2 {\rm tr} |B|^2-m_{\overline B}^2 {\rm tr} 
|\overline B|^2 - 
m_T^2 {\rm tr} |T|^2+(h'B_iT^i_j\overline B^j + h.c.).
\label{ssb2}
\end{eqnarray}
The above theory can be obtained from the dual theory with the flavour number 
$N_f=N_c+2$ by deforming the flavour number, i.e. by adding the mass term 
$m_{N_c+2}\widehat T^{N_c+2}_{N_c+2}$ into the superpotenail (\ref{spot}).
In the SUSY limit the dual gauge symmetry $SU(2)$ is broken completely and 
the nonperturbative superpotential appears due to instanton contribution, 
\begin{eqnarray}
{1 \over \Lambda^{2N_c-1}}\det \widehat T,
\label{inst}
\end{eqnarray}
where $\det \widehat T$ denotes the determinat of the 
$(N_c+1)\times (N_c+1)$ part of $\widehat T$.
We rescale $B_i=\Lambda^{N_c-1}\sqrt {m_{N_c+2}}~ q_i$ and 
$\overline B_i=\Lambda^{N_c-1}\sqrt {m_{N_c+2}}~ \overline q_i$.
Furthermore, SUSY breaking terms (\ref{ssb}) for $N_f=N_c+2$ correspond 
to those for $N_f=N_c+1$ (\ref{ssb2}) except the $(N_c+2)$-th flavour.
In particular, the trilinear coupling term $hq_iT^i_j\overline q^j$ 
corresponds to $h'B_iT^i_j\overline B^j$.

Here we consider the minimum of the potential.
The flavour symmetry allows the possibility of the further SUSY breaking term 
$h'_T\det T,$
corresponding to the nonperturbative superpotential (\ref{inst}).
Such possibility can be included in the following discussions, as shall 
be shown later.
We assume the canonical kinetic terms with the normalization factors 
$k_B$ and $k_T$ for $B$, $\overline B$ and $T$.
We have the scalar potential,
\begin{eqnarray}
V &=& \lambda^2_T \sum_{i,j} |B_i \overline B^j-(\mbox{det} '~T)_i^j|^2
+{\lambda_B^2}{\rm tr}(|\overline B T|^2
+|BT|^2) \nonumber \\
&+& 
m_B^2{\rm tr} B^\dagger B 
+ m_{\bar B}^2 {\rm tr} \overline B \overline B^\dagger  
+m_T^2 {\rm tr} T^\dagger T - (h'B_iT^i_j\overline B^j + h.c.),
\label{pot}
\end{eqnarray}
where $\lambda_T= 1/(k_T\Lambda^{2N_c-1})$, 
$\lambda_B= 1/(k_B\Lambda^{2N_c-1})$ and 
$(\det'~T)_i^j \equiv \partial \det T/\partial T^i_j$.
Here we consider the case where the SUSY breaking terms are small  
compared with $\lambda_T$ as well as with $\lambda_B$.
In Ref.\cite{soft1} the potential with $h'=0$ has been discussed. 
In this case we have only the trivial vacuum with $B=\overline B=T=0$ and 
chiral symmetry is unbroken.

The minimum of the potentail can be obtained along the diagonal direction 
(\ref{Tdiag}).
Here we consider the following direction,
\begin{eqnarray}
B_i\overline B^j -(\mbox{det} '~T)_i^j=0.
\label{flatd}
\end{eqnarray}
Along this direction, the first term in the scalar potential 
(\ref{pot}) vanishes.
For simplicity, we consider the case where 
\begin{eqnarray}
B_i=\overline B^i=B, \quad T_{(i)}=T, \mbox{ for all $i$'s,}
\end{eqnarray}
and $B$ and $T$ are real.
In this case we have the potential,
\begin{eqnarray}
{V \over N_c+1}&=& 2\lambda_B^2T^{N_c+2}-2h'T^{N_c+1} 
+ (m_B^2+m_{\overline B}^2)T^{N_c}+m_T^2T^2,
\label{pot2}
\end{eqnarray}
along the direction (\ref{flatd}).
Here the direction (\ref{flatd}) means $T \geq 0$.
If we add the SUSY breaking term $h'_T \det T$ corresponding to the 
nonperturbative superpotential, we have the extra term $h'_T T^{N_c+1}$.
That corresponds to only the shift,
\begin{eqnarray}
h' \rightarrow h'+h'_T,
\end{eqnarray}
in the scalar potential (\ref{pot2}).

For a sufficiently large value of $h'$ the nontrivial global minimum with 
$T \neq 0$ can appear.
As an example, Figs. 1 and 2 show the potentail (\ref{pot2}) for $N_c=3$.

\begin{center}
\input fig1.tex

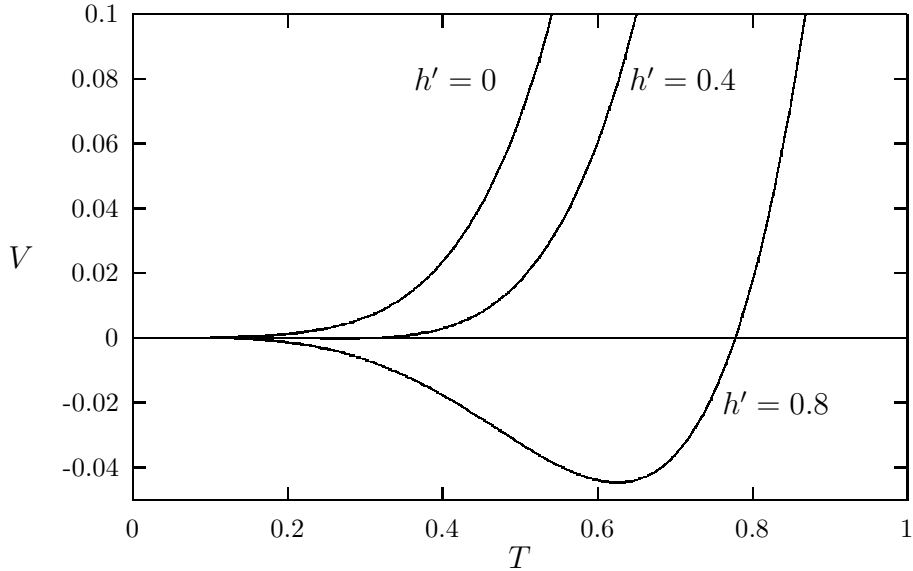
Fig.1: The scalar potential for $\lambda_B=1$, 
$m_B=m_{\overline B}=m_T=0.1$ and $h'=0, 0.4$ and 0.8.
\end{center} 
\begin{center}
\input fig2.tex

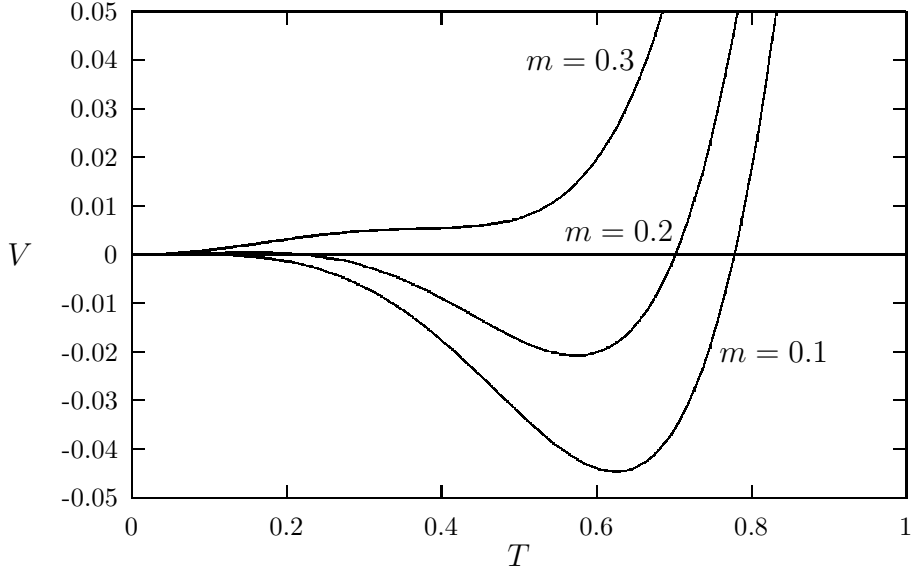
Fig.2: The scalar potential for $\lambda_B=1$, $h'=0.8$ and 
$m=m_B=m_{\overline B}=m_T=0.1, 0.2$ and 0.3.
\end{center} 

We have the nontrivial vacuum with nonvanishing $B$, $\overline B$ and 
$T$, i.e. the flavour symmetry breaking, if $h'$ is sufficiently large 
compared with the soft scalar masses.
Thus, the trilinear coupling term $h'BT\overline B$ plays an 
important role in the 
chiral symmetry breaking for the case with $N_f=N_c+1$ like the term 
$hqT\bar q$ for the case with $N_f > N_c+1$.
In both cases with $N_f=N_c+1$ and $N_f > N_c+1$, the same trilinear 
term appears to be important.

Here we give a comment on the case where SUSY breaking terms break the 
flavour symmetry $SU(N_f=N_c+1) \rightarrow SU(N_f=N_c)$.
That will be useful to understand the case with $N_f=N_c$.
As an example we consider only nonvanishing trilinear coupling term 
$h'B_{N_c+1}T^{N_c+1}_{N_c+1}\overline B^{N_c+1}$.
In the following discussion only the soft mass terms of $B_{N_c+1}$, 
$\overline B_{N_c+1}$ and $T_{(i)}$ for $i \neq N_c+1$ are relevant and 
only these soft mass terms are considered.
Now we have 
\begin{eqnarray}
V &=& \lambda^2_T \sum_{i,j} |B_i \overline B^j-(\mbox{det} '~T)_i^j|^2
+{\lambda_B^2}{\rm tr}(|\overline B T|^2
+|BT|^2) \nonumber \\
&+& 
m_{B(N_c+1)}^2|B_{N_c+1}|^2 
+ m_{\bar B(N_c+1)}^2 |\overline B^{N_c+1} |^2  
+m_T^2 \sum_{i=1}^{N_c} |T|^2 \nonumber \\
&-& (h'B_{N_c+1}T^{N_c+1}_{N_c+1}
\overline B^{N_c+1} 
+ h.c.).
\label{pot12}
\end{eqnarray}
We consider this potential for a fixed value of $T^{N_c+1}_{N_c+1}$, which 
is taken to be large enough here.
In addition, we take the case where SUSY breaking terms are sufficiently 
large.
If $m_T^2 \ll m_B^2, m_{\overline B}^2, \lambda_B^2$, we have the potential 
minimum at $B= 0$.
The relevant part is expanded near $B=0$: 
\begin{eqnarray}
V &=& \lambda^2_T \sum_{i,j} |(\mbox{det} '~T)_i^j|^2
+m_T^2 \sum_{i=1}^{N_c} |T|^2
- ( h'_T\det T+ h.c.) +\cdots.
\end{eqnarray}
In this case $T$ develops its nonvanishing finite vacuum expectation value, 
which is determined by $\lambda_T$ and $h'_T$ as well as $m_T^2$.
On the other hand, if 
$m_T^2 \gg m_B^2, m_{\overline B}^2, \lambda_B^2$, 
we have the potential minimum at $B \neq 0$ and $T=0$.

\section{$N_f \leq N_c$}

In Ref.\cite{soft1} 
the $N_f=N_c$ and $N_f < N_c$ supersymmetric QCD are broken softly 
by adding soft scalar masses.
It has been shown that in both cases with 
$N_f=N_c$ and $N_f < N_c$ we have only the nontrivial vacua leading to 
the chiral symmetry breaking.

The $N_f=N_c$ supersymmetric QCD can be described in terms of 
the baryon pair $\widehat B$ and $\widehat {\overline B}$ and 
$N_c \times N_c$ meson fields $\widehat T_i^j$.
We have the quantum constraint \cite{seiberg,rev},
\begin{eqnarray}
\widehat B\widehat {\overline B} - \det \widehat T =\Lambda^{2N_c}.
\label{const}
\end{eqnarray}
The flavour symmetry allows the SUSY breaking term $h_B B\overline B$, i.e. 
the mixing mass term of $B$ and $\overline B$.
Thus we consider the following SUSY breaking terms here,
\begin{eqnarray}
{\cal L}_{SB} &=& -m_B^2 |B|^2-m_{\overline B}^2  
|\overline B|^2 - 
m_T^2 {\rm tr} |T|^2+(h_BB\overline B + h.c.).
\label{ssb3}
\end{eqnarray} 
Following Ref.\cite{soft1} we consider the direction,
\begin{eqnarray}
B=-\overline B ={b \over \Lambda^{N_c}}, \quad T_{(i)} =t/\Lambda^2 .
\end{eqnarray}
Hence, we study the minimum of the potential, 
\begin{eqnarray}
V = 2{m'_B}^2 |b|^2   + {m'_T}^2N_c |t|^2-(h'_Bb^2 + h.c.),
\label{pot3}
\end{eqnarray} 
where ${m'_B}^2=(m_B^2+m_{\overline B}^2)/\Lambda^{2N_c}$, 
${m'_T}^2={m'_T}^2/\Lambda^{2N_c}$ and $h'_B=h_B/\Lambda^{2N_c}$, 
taking into account the constraint,
\begin{eqnarray}
t^{N_c}+b^2=1.
\end{eqnarray}
There are two stable points of the potential.
One point corresponds to $b=0$ and $|t|=1$.
The other corresponds to $t=0$ and $b=\pm 1$, where $V$ takes the value 
$V=2{m'_B}^2-2h'_B$ if $h'_B$ is real.
Thus the effect of the SUSY breaking term $h_B B\overline B$ is the 
shift $2{m'_B}^2 \rightarrow 2{m'_B}^2-2h'_B$ in determining the minimum.
Determining the global minimum among the two stable points 
depends on the ratio, 
$r \equiv (2{m'_B}^2-2h'_B)/{m'_T}^2$.
If $r$ is sufficiently large, the vacuum with $T=1$ is realized and the 
flavour symmetry is broken.
If $r$ is sufficiently small or negative, the vacuum with $b=1$ is 
realized and the baryon number symmetry is broken.
This situation is very similar to the case without the SUSY breaking term 
$h_B B\overline B$.
Even without $h_B B\overline B$, we always have a nontrivial vacuum with 
$T \neq 0$ or $b \neq 0$ for the $N_f=N_c$ case with soft scalar 
masses \cite{soft1}.
Thus, the SUSY breaking term $h_B B\overline B$ is not so important as 
the term $h' B T\overline B$ in the $N_f=N_c+1$ case.

Furthermore, there is the possibility of adding the SUSY breaking term 
$h_T \det T$.
However, that just leads to the shift $h'_B \rightarrow h'_B +h_T$ under 
the constraint (\ref{const}).

For $N_f < N_c$ the $N=1$ supersymmetric QCD has the nonperturbative 
superpotential,
\begin{eqnarray}
W= (N_c -N_f)\left( {\Lambda^{3N_c-N_f} \over \det T} \right)^{1/(N_c-N_f)}.
\end{eqnarray}
This potential has no stable point for a finite value of $T$.
However, if we add the soft SUSY breaking scalar mass term,
\begin{eqnarray}
{\cal L}_{SB} = -m_T^2 {\rm tr} |T|^2,
\label{ssb4}
\end{eqnarray} 
we have a stable vacuum for a finite value of $T$ as already shown 
in Ref.\cite{soft1}.
The soft mass terms are all the SUSY breaking terms allowed by 
the symmetries.

Up to now, we have studied the minimum of the potential in the softly 
broken $N=1$ supersymmetric QCD for different flavour numbers.
It is suggestive to compare the theory with different flavour numbers.
For $N_f > N_c+1$ and $N_f =N_c+1$, it is possible to add the trilinear 
SUSY breaking terms $hqT\overline q$ and $h'BT\overline B$.
Furthermore, in this case two types of vacua are possible.
One is the trivial vacuum with $T =0$ and the other is the 
nontrivial vacuum with $T \neq 0$ as well as $q \neq 0$ and $B \neq 0$.
Which vacuum is realized depends on the magnitude of trilinear 
couplings $h$ or $h'$.
If they are large enough, the nontrivial vacuum is realized.

On the other hand, there does not appear such trilinear terms for 
$N_f=N_c$ and $N_f < N_c$.
In this case we always have the nontrivial vacua, i.e. 
$T \neq 0$ or $B \neq 0$ for $N_f=N_c$ and  $T \neq 0$ for $N_f < N_c$.

It is interesting to go further in this comparison.
For that purpose we consider deformation of softly broken $N=1$ 
supersymmetric QCD with $N_f=N_c+1$ into the theories  with 
$N_f=N_c$ and $N_f < N_c$.
Here we add the mass term $m\widehat T^{N_c+1}_{N_c+1}$ in the superpotential 
of the $N_f=N_c+1$ theory (\ref{spot2}),
\begin{eqnarray}
W={1 \over \Lambda^{2N_c-1}}
(\widehat B_i \widehat T^i_j\widehat {\overline B}^j -\det \widehat T) 
+m\widehat T^{N_c+1}_{N_c+1}.
\label{spot3}
\end{eqnarray}
In the SUSY limit we obtain the constraint (\ref{const}) after we 
integrate out $T^{N_c+1}_{N_c+1}$ and take $m = \Lambda$.
Here we add only the soft scalar masses as SUSY breaking terms.
Then the corresponding scalar potential is obtained as 
\begin{eqnarray}
V &=& \lambda^2_T \sum_{i,j=1} |B_i \overline B^j-(\mbox{det} '~T)_i^j
+\delta^{N_c+1}_i\delta^{j}_{N_c+1}m|^2
+{\lambda_B^2}{\rm tr}(|\overline B T|^2
+|BT|^2) \nonumber \\
&+& 
m_{B(N_c+1)}^2|B_{N_c+1}|^2 
+ m_{\bar B(N_c+1)}^2 |\overline B^{N_c+1} |^2  
+m_T^2 \sum_{i=1}^{N_c} |T|^2 .
\label{pot32}
\end{eqnarray}
This scalar potential is very similar to eq.(\ref{pot12}) and 
the term $mB_{N_c+1}\overline B^{N_c+1}$ works effectively similar to 
the SUSY breaking trilinear term 
$h'B_{N_c+1}T^{N_c+1}_{N_c+1}\overline B^{N_c+1}$ in eq.(\ref{pot12}).
Thus the vacuum of the potential (\ref{pot32}), which corresponds to 
the $N_f=N_c$ flavour theory in the SUSY limit with large $m$, corresponds 
to the nontrivual vacuum of the $N_f=N_c+1$ flavour theory.
Similarly, the vacuum of the case with $N_f < N_c$ corresponds to 
the nontrivial vacuum of the case with $N_f=N_c+1$.

\section{Conclusions}

We have studied the broken $N=1$ supersymmetric QCD with all the possible 
SUSY breaking terms.
For $N_f > N_c+1$ and $N_f=N_c+1$, we have two types of vacua, namely  
one is trivial and the other is nontrivial vacuum.
Which vacuum is realized depends on the magnitude of the trilinear SUSY 
breaking terms.
If we add only the soft mass terms, we have only the 
trivial vacuum and chiral symmetry is unbroken \cite{soft1}.
Thus, the trilinear SUSY breaking terms are very important in determining the 
vacuum structure.

For $N_f = N_c$ and $N_f < N_c$, we can not have the trilinear SUSY breaking 
term and we always have the nontrivial vacua.

Reduction of the flavour number is realized by adding a suitable trilinear 
SUSY breaking term.
If we have still trilinear SUSY breaking terms for massless modes, i.e. 
for $N_f \geq N_c+1$, 
we have the possibility of having the trivial vacuum only for the massless 
modes.
Otherwise, i.e. for $N_f \leq N_c$, we have only nontrivial vacua.
The vacua with $N_f = N_c$ and $N_f < N_c$ correspond to 
the nontrivial vacuum of the $N_f=N_c+1$ theory with a large 
trilinear coupling for the $(N_c+1)$-th flavour.

\section*{Acknowledgments}  
This work was partially supported by the Academy of Finland under 
Project no. 37599.


\begin{thebibliography}{99}

\bibitem{seiberg}
N. Seiberg, Phys. Rev. {\bf D49} (1994) 6857;
Nucl. Phys. {\bf B435} (1995) 129.

\bibitem{rev}
For reviews see, e.g.
K.~Intriligator and N.¨Seiberg, 
Nucl. Phys. Proc. Suppl. {\bf 45BC} (1996) 1;\\
L.~\'Alvarez-Gaum\'e and S.F.~Hassan, CERN-TH/96-371, hep-th/9701069;\\
M.~Shifman, TPI-MINN-97/09-T, hep-th/9704114.

\bibitem{soft1}
O.~Aharony, J.~Sonnenschein, M.E.~Peskin and S.~Yankielowicz, 
Phys. Rev. {\bf D52} (1995) 6157.

\bibitem{soft2}
E.~D'Hoker, Y.~Mimura and N.~Sakai, 
Phys. Rev. {\bf D54} (1996) 7724.

\bibitem{soft3}
N.~Evans, S.D.H.~Hsu and M.~Schwets, 
Phys. Lett. {\bf B404} (1997) 77;\\
H.C.~Cheng and Y.~Shadmi, Fermilab-Pub-97/420-T,
hep-th/9801146;\\
S.P.~Martin and J.~D.~Wells, SLAC-PUB-7739, hep-th/9801157.

\bibitem{soft4}
N.~Evans, S.D.H.~Hsu, M.~Schwets and S.B.~Selipsky,
Nucl. Phys. {\bf B456} (1995) 205;\\
N.~Evans, S.D.H.~Hsu and M.~Schwets, 
Nucl. Phys. {\bf B484} (1997) 124;\\
L.~\'Alvarez-Gaum\'e, J.~Distler, C.~Kounnas and M.~Mari\~no,
Int. J. Mod. Phys. {\bf 11} (1996) 4745;\\
L.~\'Alvarez-Gaum\'e and M.~Mari\~no,
Int. J. Mod. Phys. {\bf 12} (1997) 975;\\
K.~Konishi, Phys. Lett. {\bf B392} (1997) 101;\\
L.~\'Alvarez-Gaum\'e, M.~Mari\~no and F.~Zamora,
CERN-TH/97-37, hep-th/9703072; CERN-TH/97-144, hep-th/9707017.


\bibitem{mqcd}
A.~Brandhuber, J.~Sonnenschein, S.~Theisen and S.~Yankielowicz, 
Nucl. Phys. {\bf B502} (1997) 125;\\
A.~Hanany, M.J.~Strassler and A.~Zaffaroni, 
Nucl. Phys. {\bf B513} (1998) 87;\\
N.~Evans and M.~Schwetz, Nucl. Phys. {\bf B522} (1998) 69;\\
J.L.F.~Barb\'on and A.~Pasquinucci, 
Phys. Lett. {\bf B421} (1998) 131; 
Mod. Phys. Lett. {\bf A13} (1998) 1453;\\
N.~Evans, BUHEP-98-03, hep-th/9801159; BUHEP-98-08, hep-th/9804097.

\bibitem{soft5}
N.~Arkani-Hamed and R.~Rattazzi, SLAC-PUB-7777, hep-th/9804068.


\bibitem{soft6}
M.~Chaichian, W.F.~Chen and T.~Kobayashi, 
Phys. Lett. {\bf B432} (1998) 120.




\end{thebibliography}
\end{document}